\documentclass[pdflatex,sn-mathphys-num]{sn-jnl}

\usepackage{graphicx}
\usepackage{amsmath,amssymb,amsfonts}
\usepackage{amsthm}
\usepackage{mathtools}
\usepackage{booktabs}
\usepackage{xcolor}
\hypersetup{pdftitle={Hidden convexity of the Eliashberg condensation energy},pdfauthor={Davis Rens},pdfkeywords={Eliashberg theory, superconductivity, spin chain, convexity, uniqueness, phase transition, formal verification},bookmarksdepth=2}
\renewcommand{\pacsbullet}{\hbox{,\hskip2.5pt}}

\theoremstyle{thmstyleone}
\newtheorem{theorem}{Theorem}
\newtheorem{lemma}[theorem]{Lemma}
\newtheorem{proposition}[theorem]{Proposition}
\newtheorem{corollary}[theorem]{Corollary}
\theoremstyle{thmstyletwo}
\newtheorem{remark}[theorem]{Remark}
\theoremstyle{thmstylethree}
\newtheorem{definition}[theorem]{Definition}

\newcommand{\ip}[2]{\langle #1,#2\rangle}
\newcommand{\NN}{\mathbb{N}}
\newcommand{\RR}{\mathbb{R}}

\newcommand{\Rset}{\mathcal{R}}
\newcommand{\Sp}{\operatorname{sup\,spec}}

\begin{document}

\title[Hidden convexity of the Eliashberg condensation energy]{Hidden convexity of the Eliashberg condensation energy: existence, uniqueness and continuity of the superconducting state}

\author*{\fnm{Davis} \sur{Rens}}\email{davis@supermatics.io}

\affil*{\orgname{SuperMatics Inc.}, \orgaddress{\city{Berkeley}, \state{CA}, \country{USA}}}

\abstract{\unboldmath Kiessling, Altshuler and Yuzbashyan conjectured the thermodynamic transition picture for the isotropic Eliashberg spin-chain functional with vanishing Coulomb pseudopotential: the normal state is the unique minimizer above and at a critical temperature, the superconducting minimizer below it is unique up to gauge, and the transition is continuous. We establish the nonlinear part of this picture. In the variables $\rho_n=\sin^2\theta_n$, the condensation energy restricted to the first quadrant is strictly convex, and folding any admissible chain into that quadrant cannot increase its energy. For every positive, non-increasing interaction with power-law decay, hidden convexity and a compactness estimate give existence and uniqueness of the minimizer up to gauge. The normal state minimizes the energy exactly when its quadratic form is non-negative, including marginal stability. Among admissible chains, a positive solution of the nonlinear Eliashberg equations exists exactly when the normal state is linearly unstable and is then unique. For temperature-dependent families with pointwise continuity and locally uniform decay bounds, the minimizing density is continuous in a weighted $\ell^1$ norm. Together with uniqueness of the linear-stability temperature, these results prove the thermodynamic conjectures for the $\gamma$-model, Einstein phonons and general phonon spectra; the phonon applications use separately established spectral input. Finite-dimensional algebraic and convex-analytic components are machine-checked in Lean. The proofs were obtained independently of concurrent work that reaches overlapping conclusions.}

\keywords{Eliashberg theory, superconductivity, classical spin chain, convexity, uniqueness of minimizers, phase transitions}

\pacs[MSC Classification]{82D55, 49J45, 49K27, 47A75, 68V15}

\maketitle

\section{Introduction}\label{sec:intro}

\subsection{Background}

Eliashberg theory describes phonon-mediated superconductivity through a gap function $\Delta_n$ and a mass renormalization $Z_n$ on the fermionic Matsubara frequencies $\omega_n=\pi T(2n+1)$~\cite{Eliashberg1960,AllenMitrovic1982,Carbotte1990,Marsiglio2020}. Yuzbashyan and Altshuler~\cite{YuzbashyanAltshuler2022SpinChain} showed that its free energy is that of a classical Heisenberg chain in which each Matsubara frequency carries a unit spin, and that the Eliashberg equations are the conditions for the spins to be in equilibrium. Kiessling, Altshuler and Yuzbashyan (KAY) took this formulation as the starting point for a mathematical study of the theory with vanishing Coulomb pseudopotential, in three papers on the $\gamma$-model~\cite{KAY2025I}, on dispersive phonons~\cite{KAY2025II} and on Einstein phonons~\cite{KAY2025III}. In each of them the transition is described by a thermodynamic narrative, which they stated as a conjecture~\cite[Conj.~1]{KAY2025I}, \cite[Conj.~1]{KAY2025II}, \cite[Conj.~2]{KAY2025III}. For the $\gamma$-model it reads:

\begin{quote}
\emph{There is a critical temperature $T_c(\gamma)>0$ such that for temperatures $T\ge T_c$, the spin chain of the normal state is the unique minimizer of $H(S|N)$, whereas when $T<T_c$ a spin chain $S\ne N$ for a superconducting phase minimizes $H(S|N)$ uniquely up to an irrelevant gauge transformation (fixing of an overall phase). Moreover, the phase transition at $T_c$ from normal to superconductivity is continuous.} \hfill\cite[Conj.~1]{KAY2025I}
\end{quote}

KAY studied the linearization of the theory about the normal state and located the temperature at which the normal state loses linear stability. In the $\gamma$-model paper they assumed that minimizers exist, ``leaving a rigorous inquiry into the existence and uniqueness of minimizers to a future work''~\cite[Sect.~2]{KAY2025I}, and in the paper on dispersive phonons they wrote that the normal state minimizes the condensation energy ``only for sufficiently large $T$, as we will prove in a separate paper''~\cite[Sect.~4]{KAY2025II}. Kondratev's proof of KAY's bound on the stability temperature likewise states that identifying it with the thermodynamic transition is an assumption, not a theorem proved there~\cite{Kondratev2026}; the argument is an archived preprint, not a peer-reviewed publication. Elezaby and Abanov derived rigorous upper and lower bounds on the transition temperature of the $\gamma$-model from the Hessian of the free energy~\cite{ElezabyAbanov2026}, again an analysis of the linearized problem. Yuzbashyan and Altshuler showed that a minimizer has all its angles in $[0,\pi/2]$ (as recalled in~\cite[Sect.~4]{KAY2025II}) and that there is no continuous family of stationary points connected to the global minimum, and argued from symmetry that the global minimum is unique up to the overall phase~\cite{YuzbashyanAltshuler2022SpinChain}. For the $\gamma$-model with $\gamma>3$, in the limit $T\to0$ of extreme retardation, Yuzbashyan, Kiessling and Altshuler reduced the problem to a universal boundary-value problem for the angle and proved that the global minimum of the free energy is non-degenerate~\cite{YKA2022}; their proof, by a monotonicity argument for $\theta''=x\sin\theta$~\cite[App.~A]{YKA2022}, concerns that limiting local functional. In the peer-reviewed literature, existence, uniqueness and the relation between the nonlinear and the linear problem for the discrete chain at positive temperature were open; concurrent work is discussed in Section~\ref{sec:concurrent}. The $\gamma$-model at $T=0$ is known to have an infinite tower of solutions with $n$ sign changes, of which the sign-preserving one is argued to be the global minimum and the others saddle points~\cite{AbanovChubukov2020,ZhangWuAbanovChubukov2022}; our results concern $T>0$ and are consistent with this picture.

The corresponding questions for BCS theory have a long history. Existence and uniqueness of the positive solution of the BCS gap equation were proved under various hypotheses~\cite{Odeh1964,BillardFano1968,Vansevenant1985,Yang1991,Du1995,McLeodYang2000}, Hainzl, Hamza, Seiringer and Solovej proved that the normal state is the global minimizer of the BCS functional if and only if a linear operator has no negative spectrum~\cite{HHSS2008}, and Watanabe gave an operator-theoretical proof that the transition in the BCS--Bogoliubov model is of second order~\cite{Watanabe2020}. Closest to our method, Hainzl and Seiringer proved that the zero-temperature BCS minimizer is unique, for radial potentials $V\in L^1(\RR^3)\cap L^{3/2}(\RR^3)$ with $\hat V\le0$ and $\hat V(0)<0$, using the strict convexity of $t\mapsto1-\sqrt{1-4t}$ in the squared pair amplitude $t=|\hat\alpha|^2$; the unique representative is the one whose Fourier transform is non-negative~\cite[Sect.~5.2]{HainzlSeiringer2008}. Our function $g(r)=1-\sqrt{1-r}$ in $r=\sin^2\theta$ plays the same role. The step that is new here is an exact regrouping of the two frequency-transfer sums: written in $x_n,c_n$, they become a complete square in $c_n-c_m$ with a non-negative coefficient plus a negative geometric-mean term in the $x_n$, and it is this regrouping, not frequency nonlocality as such, that exhibits the contributions that are convex in $\rho$. Each sum separately is not convex in the angles. Rigorous results on Eliashberg theory itself also include an obstruction to odd-frequency pairing~\cite{LangmannEtAl2022}, an archived preprint with the stable identifier \texttt{10.48550/arXiv.2207.01825}.

\subsection{Results}\label{sec:overview}

Write $\Theta=(\theta_n)_{n\in\NN_0}$ for the vector of spin angles, $\theta_n$ being the angle of the $n$-th spin from the normal-state direction, so that $\tan\theta_n=\Delta_n/\omega_n$~\cite{YuzbashyanAltshuler2022SpinChain}. Our main observation is that the condensation energy, restricted to angles in $[0,\pi/2]$ and written in the variables
\[
\rho_n:=\sin^2\theta_n=\frac{\Delta_n^2}{\omega_n^2+\Delta_n^2},
\]
is a strictly convex functional, and that every minimizer lies in this quadrant up to gauge. We prove the following for every pair interaction $\lambda(k)$ that is positive, non-increasing in the frequency transfer $k$ and bounded by $Ck^{-\gamma}$ for some $C,\gamma>0$ (Definition~\ref{def:kernel}); this covers the $\gamma$-model for every $\gamma>0$, Einstein phonons, and every phonon spectrum with finite second moment.
\begin{enumerate}
\item \emph{Existence and uniqueness} (Theorem~\ref{thm:exist}). The condensation energy has a minimizer, and the minimizer is unique up to the gauge $\Theta\mapsto-\Theta$.
\item \emph{Linear and nonlinear stability coincide} (Theorem~\ref{thm:stab}). The normal state is the minimizer if and only if it is linearly stable, i.e.\ if and only if KAY's quadratic form is non-negative. At the stability boundary the normal state is still the unique minimizer.
\item \emph{The positive Eliashberg solution} (Theorem~\ref{thm:EL}). Among admissible chains, the nonlinear Eliashberg equations have a solution with all $\Delta_n>0$ if and only if the normal state is linearly unstable. It is then unique, and it is the minimizer.
\item \emph{Continuity} (Theorem~\ref{thm:cont}). The minimizer depends continuously on temperature in a weighted $\ell^1$ norm, so the order parameter vanishes continuously as the stability boundary is approached.
\end{enumerate}
The critical temperature in KAY's conjectures is the temperature at which the normal state loses linear stability. For the $\gamma$-model the stability set is an interval $[T_c,\infty)$ by scaling, and Results~1--4 prove KAY's Conjecture for the $\gamma$-model outright. For Einstein phonons and general spectra, the linear-stability temperature is unique for every coupling by the strict monotonicity in $T$ of the top eigenvalue of KAY's stability operator. This monotonicity is proved in our companion paper, \emph{Sharp bounds on the Eliashberg critical temperature: a comparison principle, convexity in the spectrum, and the Kiessling--Altshuler--Yuzbashyan conjectures} (D.~Rens, 2026, submitted), and, independently, in the archived preprint of Kondratev~\cite{Kondratev2026}, an online document with the stable identifier \texttt{10.5281/zenodo.22848039} that has not been peer reviewed; KAY proved uniqueness of the crossing above a spectrum-dependent coupling threshold, using monotonicity on a restricted temperature range. With it, Results~1--4 prove KAY's conjectures for dispersive and Einstein phonons (Corollary~\ref{cor:kay}). The thermodynamic and linear-stability critical temperatures therefore coincide in this model.

\subsection{Note on concurrent work}\label{sec:concurrent}

The results of this paper were obtained independently; the proofs were complete on 10 November 2026. Since then two works with overlapping results have been made public. Neither has been peer reviewed, and we have not checked their proofs; we describe them as their authors state them.

Siraj released two manuscripts with a Lean~4 formalization on 13 November 2026~\cite{Siraj2026} (we cite version 1.0.2 of 14 November). At $\mu^*=0$ they prove KAY's thermodynamic conjectures by the mechanism used here: the condensation energy is written as its quadratic form plus non-negative squares, and in the variables $\rho_n=\sin^2\theta_n$ on the first quadrant it becomes strictly convex. They go further in several directions: the global minimizer is unique among three-dimensional spin configurations, without assuming time-reversal symmetry or frequency parity, for positive, summable, non-increasing kernels; the condensation energy is bounded below by $-C(T_c-T)^2$ as $T\uparrow T_c$; and the minimization is transferred to the Yuzbashyan--Altshuler free energy of the self-energies. On 16 November 2026 Generalised Superintelligence published a white paper~\cite{GSI2026} stating that, without spin flips, the nonlinear isotropic Eliashberg equations have no nonzero real even-frequency solution at or above the linear-instability temperature, also with a symmetric Coulomb term, and that at $\mu^*=0$ exactly one positive solution exists below it, which minimizes the condensation energy up to sign; it also excludes complex even-frequency solutions wherever the linear stability form is non-negative.

The present paper gives a self-contained proof for the planar chain, in which uniqueness, the coincidence of linear and nonlinear stability and the classification of positive solutions all follow from one rearrangement identity (Lemma~\ref{lem:identity}) and strict convexity (Proposition~\ref{prop:convex}). It applies to every positive, non-increasing interaction with power-law decay, summable or not, and proves continuity of the minimizer in the weighted $\ell^1$ norm of Theorem~\ref{thm:cont}, which controls every $\Delta_n$. The variational problem for the condensation energy $H(S|N)=4\pi^2T\,K(\Theta)$, the free-energy difference between a chain and the normal state, is solved over all admissible real chains, including spin-flipped ($\cos\theta_n<0$) and sign-changing ones. At and above $T_c$, every non-normal admissible chain has strictly positive condensation energy (Theorem~\ref{thm:stab}(b)). So no first-order transition into a real-gap state can occur, there is no positive-gap stationary point (Theorem~\ref{thm:EL}), and the thermodynamic transition temperature equals the linear-stability temperature. We do not treat non-planar spin configurations; for these see~\cite{Siraj2026}.

\subsection{The idea of the proof}

The condensation energy is a sum of a Zeeman-like term $\sum_n(2n+1)(1-\cos\theta_n)$ and of pair terms $\lambda(|n-m|)(1-\cos(\theta_n-\theta_m))$ and $-\lambda(n+m+1)(1-\cos(\theta_n+\theta_m))$, the last pair term arising from the spins at negative frequencies. With $x_n=\sin\theta_n$ and $c_n=\cos\theta_n$, the identity $1-c_nc_m=\tfrac12(x_n^2+x_m^2)+\tfrac12(c_n-c_m)^2$ rearranges it exactly (Lemma~\ref{lem:identity}) into
\[
\sum_n(2n+1)(1-c_n)+\sum_n\Lambda_nx_n^2+\frac14\sum_{n\ne m}D_{nm}(c_n-c_m)^2-\frac12\sum_{n,m}W_{nm}x_nx_m,
\]
with $\Lambda_n=\sum_{k=1}^n\lambda(k)\ge0$, $D_{nm}=\lambda(|n-m|)-\lambda(n+m+1)\ge0$ and $W_{nm}=[n\ne m]\,\lambda(|n-m|)+\lambda(n+m+1)>0$, where $[\cdot]$ denotes the indicator of a statement. In this form the gauge reduction is immediate: replacing $(x_n,c_n)$ by $(|x_n|,|c_n|)$ does not increase any term. In the quadrant $x_n,c_n\ge0$ write $x_n=\sqrt{\rho_n}$ and $c_n=\sqrt{1-\rho_n}$. Then $1-\sqrt{1-\rho_n}$ is strictly convex, $x_n^2=\rho_n$ is linear, $-\sqrt{\rho_n\rho_m}$ is convex because the geometric mean is concave, and $(\sqrt{1-\rho_n}-\sqrt{1-\rho_m})^2$ is convex for the same reason. Strict convexity gives uniqueness. The same identity shows that the functional is its quadratic part plus two non-negative remainders, which gives the stability criterion. Existence follows from a Schur-test estimate showing that the coupling term is a compact perturbation of the Zeeman term. That every positive solution of the Eliashberg equations is the minimizer is the infinite-dimensional version of the statement that a critical point of a convex function is a global minimizer.

The convexity is hidden in two senses: the functional is not convex in the angles, and it is not convex in the gap. With Coulomb repulsion, the coefficient signs used in this convexity proof are no longer automatic; we do not treat the general case $\mu^*>0$ (Section~\ref{sec:scope}).

\subsection{Verification and organization}

The algebraic and convex-analytic core, for every finite truncation of the chain, is formalized in Lean~4 with Mathlib (Section~\ref{sec:verify}). The infinite-dimensional steps (compactness, existence, passage to the limit) are proved on paper. Section~\ref{sec:numerics} reports selected finite-truncation consistency checks for seven representative interactions, at up to 400 Matsubara frequencies, compared with fixed-point solutions of the Eliashberg equations; they are illustrations and not proofs.

Section~\ref{sec:setting} sets up the functional. Section~\ref{sec:results} states the results. Sections~\ref{sec:identities}--\ref{sec:proofs} contain the proofs, Section~\ref{sec:verify} the verification and Section~\ref{sec:scope} the scope and open problems.

\section{The condensation energy}\label{sec:setting}

\subsection{Interactions}

\begin{definition}\label{def:kernel}
An \emph{interaction} is a sequence $\lambda=(\lambda(k))_{k\ge1}$ of positive numbers that is non-increasing in $k$ and satisfies $\lambda(k)\le Ck^{-\gamma}$ for all $k\ge1$, for some constants $C>0$ and $\gamma>0$.
\end{definition}

The value $\lambda(0)$ (the coupling constant for phonons, infinite for the $\gamma$-model) never enters below. The three families studied by KAY are interactions:
\begin{itemize}
\item \emph{General phonon spectra}~\cite{KAY2025II}. For a probability measure $P$ on $(0,\infty)$ with $\langle\omega^2\rangle=\int\omega^2P(d\omega)<\infty$, a coupling $\lambda>0$ and $T>0$, set $\lambda(k)=\lambda a_k(P,T)$ with $a_k(P,T)=\int\omega^2(\omega^2+(2\pi Tk)^2)^{-1}P(d\omega)$. This equals $2\int\alpha^2F(\omega)\,\omega(\omega^2+(2\pi Tk)^2)^{-1}d\omega$ for $\lambda P(d\omega)=2\omega^{-1}\alpha^2F(\omega)d\omega$. It is positive and decreasing, and $\lambda(k)\le\lambda\langle\omega^2\rangle(2\pi T)^{-2}k^{-2}$. KAY's admissible spectra have finite second moment.
\item \emph{Einstein phonons}~\cite{KAY2025III}: $P=\delta_\Omega$, $\lambda(k)=\lambda\varpi^2/(\varpi^2+k^2)$ with $\varpi=\Omega/2\pi T$.
\item \emph{The $\gamma$-model}~\cite{KAY2025I}: $\lambda(k)=\beta^\gamma k^{-\gamma}$ with $\beta=g/2\pi T$, for any $\gamma>0$. For $\gamma\le1$ this interaction is not summable.
\end{itemize}

Write $\NN_0=\{0,1,2,\dots\}$, $w_n=2n+1$, and for $n,m\in\NN_0$
\begin{equation}\label{eq:coeffs}
\begin{gathered}
\Lambda_n:=\sum_{k=1}^n\lambda(k),\qquad D_{nm}:=\lambda(|n-m|)-\lambda(n+m+1)\quad(n\ne m),\\
W_{nm}:=[n\ne m]\,\lambda(|n-m|)+\lambda(n+m+1),
\end{gathered}
\end{equation}
where $[\cdot]$ is the indicator of a statement: the term $[n\ne m]\,\lambda(|n-m|)$ is omitted when $n=m$, it is not an evaluation of $\lambda(0)$, and indeed $\lambda(0)$ (the phonon coupling constant, infinite for the $\gamma$-model) enters nowhere. Thus $\Lambda_0=0\le\Lambda_n$, $D_{nm}=D_{mn}\ge0$ because $\lambda$ is non-increasing and $|n-m|<n+m+1$, and $W_{nm}=W_{mn}>0$.

\begin{lemma}\label{lem:Dsum}
For every $n\in\NN_0$ the series $\sum_{m\ne n}D_{nm}$ converges, and
\[
\sum_{m\ne n}D_{nm}=2\Lambda_n+\lambda(2n+1)\le w_n\lambda(1).
\]
\end{lemma}

\begin{proof}
For $M>n$, splitting $m<n$ and $n<m\le M$,
\[
\sum_{m\le M,\,m\ne n}\lambda(|n-m|)=\Lambda_n+\sum_{k=1}^{M-n}\lambda(k),\qquad\sum_{m\le M,\,m\ne n}\lambda(n+m+1)=\sum_{k=n+1}^{n+M+1}\lambda(k)-\lambda(2n+1).
\]
The difference is $2\Lambda_n+\lambda(2n+1)-\sum_{k=M-n+1}^{M+n+1}\lambda(k)$, and the last sum has $2n+1$ terms, each at most $\lambda(M-n+1)\to0$. The terms $D_{nm}$ are non-negative, so the series converges to the limit of the partial sums. The bound uses $\lambda(k)\le\lambda(1)$.
\end{proof}

\subsection{The functional}

The angles $\Theta=(\theta_n)_{n\in\NN_0}$, with $\theta_n\in\RR/2\pi\mathbb Z$, are measured from the normal-state direction $N_0$, of the effective spin chain $S\in(S^1)^{\NN_0}$ of~\cite{KAY2025II}; the spins at negative frequencies are determined by the symmetry relation of~\cite{KAY2025II}. KAY write the condensation energy as $H(S|N)=4\pi^2T\,K(\Theta)$ with~\cite[eq.~(34)]{KAY2025II}, \cite[eq.~(35)]{KAY2025I}, \cite[Sect.~3]{KAY2025III}
\begin{multline}\label{eq:K}
K(\Theta)=\sum_nw_n(1-\cos\theta_n)+\frac12\sum_{n\ne m}\lambda(|n-m|)\big(1-\cos(\theta_n-\theta_m)\big)\\
-\frac12\sum_{n,m}\lambda(n+m+1)\big(1-\cos(\theta_n+\theta_m)\big).
\end{multline}
For the $\gamma$-model with $\gamma\le1$ the two double sums can diverge separately, and KAY~\cite[eq.~(35)]{KAY2025I} write them as one double sum over $n\ne m$ of the difference of the two summands, plus the diagonal terms $-\frac12\lambda(2n+1)(1-\cos2\theta_n)$ of the second sum. We use this grouped form throughout; for summable $\lambda$ it equals~\eqref{eq:K} whenever the latter converges absolutely. KAY call $\Theta$ admissible if the series, so written, converge absolutely~\cite[Sect.~2]{KAY2025II}.

\begin{definition}\label{def:adm}
$\Theta$ is \emph{admissible} if $\sum_nw_n(1-\cos\theta_n)<\infty$. We write $x_n=\sin\theta_n$, $c_n=\cos\theta_n$ and $\rho_n=x_n^2$.
\end{definition}

For admissible $\Theta$ one has $\sum_nw_n\rho_n<\infty$, since $x_n^2=(1-c_n)(1+c_n)\le2(1-c_n)$.

\begin{lemma}[Compactness of the coupling]\label{lem:schur}
Let $\lambda$ be an interaction. The matrix $M_{nm}:=W_{nm}/\sqrt{w_nw_m}$ defines a bounded, compact, self-adjoint operator $M$ on $\ell^2(\NN_0)$ with non-negative entries. If $P_N$ denotes the orthogonal projection onto the span of the first $N$ unit vectors, then $\|M-P_NMP_N\|\le\tau_N$, where $\tau_N\to0$ depends only on $C$ and $\gamma$.
\end{lemma}

\begin{proof}
Put $\bar\lambda(k):=C\max(k,1)^{-\gamma}$ for $k\ge0$. Then $W_{nm}\le2\bar\lambda(|n-m|)$ for all $n,m$: for $n\ne m$ because $\lambda(n+m+1)\le\lambda(|n-m|)$, and for $n=m$ because $W_{nn}=\lambda(2n+1)\le C$. With $p_n:=w_n^{-1/2}$,
\[
\sum_m\frac{2\bar\lambda(|n-m|)}{\sqrt{w_nw_m}}\,p_m=2S_n\,p_n,\qquad S_n:=\sum_{m\ge0}\frac{\bar\lambda(|n-m|)}{w_m}.
\]
We show that $S_n$ is finite and $S_n\to0$. At $n=0$ one has $S_0=\sum_{m\ge0}\bar\lambda(m)/w_m\le C+C\sum_{m\ge1}m^{-1-\gamma}<\infty$, so only $n\ge1$ needs estimates. For $n\ge1$ and $m\le n/2$ one has $|n-m|\ge n/2$, and these terms contribute at most $C\max(n/2,1)^{-\gamma}\sum_{m\le n/2}w_m^{-1}\le C\max(n/2,1)^{-\gamma}(1+\tfrac12\log(n+1))$. For $n/2<m\le2n$ one has $w_m>n$, and these terms contribute at most $n^{-1}\sum_{|j|\le2n}\bar\lambda(|j|)\le n^{-1}C(1+2\sum_{j=1}^{2n}j^{-\gamma})$, which tends to $0$ because $\sum_{j\le2n}j^{-\gamma}=o(n)$. For $m>2n$ one has $m-n>m/2$, and these terms contribute at most $\sum_{m>2n}C(m/2)^{-\gamma}/(2m)\le 2^{\gamma}C\sum_{m>2n}m^{-1-\gamma}\to0$. Hence $\sigma_N:=\sup_{n\ge N}S_n$ is finite for every $N$ and tends to $0$.

The Schur test states that a matrix $K$ with $\sum_m|K_{nm}|p_m\le\alpha p_n$ for all $n$ and $\sum_n|K_{nm}|p_n\le\beta p_m$ for all $m$, with $p>0$, has $\|K\|\le\sqrt{\alpha\beta}$: for $u,v\in\ell^2$ and $t>0$, $|u_n||v_m|\le\frac12\big(t\,u_n^2p_m/p_n+t^{-1}v_m^2p_n/p_m\big)$, so $|\ip{u}{Kv}|\le\frac12(t\alpha\|u\|^2+t^{-1}\beta\|v\|^2)$, and one optimizes over $t$. Since $0\le M_{nm}\le2\bar\lambda(|n-m|)/\sqrt{w_nw_m}$ and the bound is symmetric, $\|M\|\le2\sigma_0$. The matrix $(1-P_N)M$ has row sums $\le2\sigma_Np_n$ and column sums $\le2\sigma_0p_m$, so $\|(1-P_N)M\|\le2\sqrt{\sigma_N\sigma_0}$; the same holds for $P_NM(1-P_N)=(\,(1-P_N)MP_N)^{\mathsf T}$. Hence $\|M-P_NMP_N\|\le4\sqrt{\sigma_N\sigma_0}=:\tau_N\to0$. $M$ is a norm limit of finite-rank operators, hence compact.
\end{proof}

\begin{lemma}[Well-definedness]\label{lem:welldef}
Let $\lambda$ be an interaction and $\Theta$ admissible. Then every series in~\eqref{eq:K} (in the grouped form) and in the identities of Section~\ref{sec:identities} converges absolutely, and $K(\Theta)$ is finite. Conversely, if KAY's rewritten series converges absolutely, then $\Theta$ is admissible. Hence the admissible class of Definition~\ref{def:adm} is KAY's.
\end{lemma}

\begin{proof}
For $n\ne m$, by the addition theorem,
\begin{equation}\label{eq:pair}
\lambda(|n-m|)\big(1-\cos(\theta_n-\theta_m)\big)-\lambda(n+m+1)\big(1-\cos(\theta_n+\theta_m)\big)=D_{nm}(1-c_nc_m)-W_{nm}x_nx_m .
\end{equation}
Put $a_n:=1-c_n\in[0,2]$. Then $0\le1-c_nc_m=a_n+a_m-a_na_m\le a_n+a_m$, so the first term on the right of~\eqref{eq:pair} lies in $[0,D_{nm}(a_n+a_m)]$, and its double sum over $n\ne m$ is at most $2\sum_na_n\sum_{m\ne n}D_{nm}\le2\lambda(1)\sum_nw_na_n$ by Lemma~\ref{lem:Dsum}. Likewise $(c_n-c_m)^2=(a_n-a_m)^2\le2(a_n+a_m)$ and $\rho_n=a_n(2-a_n)\le2a_n$, so the series $\sum_{n\ne m}D_{nm}(c_n-c_m)^2$ and $\sum_n\Lambda_n\rho_n$ converge absolutely as well. For the second term, $\sum_{n,m}W_{nm}|x_n||x_m|=\ip{\xi}{M\xi}\le\|M\|\,\|\xi\|^2$ with $\xi_n:=\sqrt{w_n}|x_n|$, and $\|\xi\|^2=\sum_nw_n\rho_n<\infty$. The diagonal terms satisfy $\lambda(2n+1)(1-\cos2\theta_n)=2\lambda(2n+1)\rho_n$. So the grouped series converges absolutely. When $\lambda$ is summable, each of the two double sums in~\eqref{eq:K} converges absolutely as well: $1-\cos(a\mp b)\le2(1-\cos a)+2(1-\cos b)$, $\sum_{m\ne n}\lambda(|n-m|)\le2\sum_{k\ge1}\lambda(k)$ and $\sum_m\lambda(n+m+1)\le\sum_{k\ge1}\lambda(k)$. The series in Section~\ref{sec:identities} are dominated by the same bounds. Conversely, absolute convergence of KAY's rewritten series includes that of the single sum $\sum_nb_n$, $b_n:=u_n-d_n$, with $u_n:=w_n(1-c_n)$ and $d_n:=\frac12\lambda(2n+1)(1-\cos2\theta_n)$. Since $0\le d_n\le2\lambda(2n+1)(1-c_n)$ and $\lambda(2n+1)\to0$, there is $N$ with $4\lambda(2n+1)\le w_n$ for $n\ge N$; then $d_n\le\frac12u_n$, so $u_n\le2b_n$ for $n\ge N$, and $\sum_nu_n=\sum_nw_n(1-c_n)<\infty$.
\end{proof}

The normal state $\Theta=0$ has $K(0)=0$. The functional is invariant under $\Theta\mapsto-\Theta$, the reflection of every spin about the normal-state direction $N_0$ (equivalently $K_0\mapsto-K_0$ in the notation of~\cite{KAY2025I}). For the effective chain in $(S^1)^{\NN_0}$ this is the ``irrelevant gauge transformation (fixing of an overall phase)'' of KAY's conjecture~\cite[Sect.~2]{KAY2025I}. The two representatives are the residual sign choices after selecting a real pairing plane; restoring the common complex phase produces the usual phase orbit. Our minimization statements concern the effective $S^1$ chains specified above.

\subsection{The quadratic form and the stability operator}

The second-order expansion of $K$ about the normal state is KAY's quadratic form~\cite[eq.~(36)]{KAY2025II},
\begin{equation}\label{eq:Q}
Q(v):=\frac12\sum_nw_nv_n^2+\sum_n\Lambda_nv_n^2-\frac12\sum_{n,m}W_{nm}v_nv_m,
\end{equation}
defined for real sequences $v$ with $\sum_nw_nv_n^2<\infty$; the middle sum is then finite because $\Lambda_n\le n\lambda(1)<w_n\lambda(1)$. With $\eta_n=\sqrt{w_n}v_n$, $Q(v)=\frac12\|\eta\|^2+\sum_n\Lambda_nw_n^{-1}\eta_n^2-\frac12\ip{\eta}{M\eta}$. For an interaction $\lambda(k)=\lambda a_k$ this is
\begin{equation}\label{eq:QK}
\begin{gathered}
Q(v)=\tfrac12\ip{\eta}{(I-\lambda\mathsf K[a])\eta},\\
\mathsf K[a]_{nm}=\frac{[n\ne m]a_{|n-m|}+a_{n+m+1}}{\sqrt{w_nw_m}}-\delta_{nm}\frac{2A_n}{w_n},\qquad A_n=\sum_{k=1}^na_k,
\end{gathered}
\end{equation}
which is the stability operator $K(P,T)$ of KAY~\cite[eqs.~(39)--(42)]{KAY2025II} in the normalization of the companion manuscript described in Section~\ref{sec:overview}. Since $\Lambda_n/w_n\le\lambda(1)$, $\mathsf K[a]$ is bounded, and it is compact by Lemma~\ref{lem:schur} together with $\Lambda_n/w_n\to0$; the latter holds because $\Lambda_n/w_n=(n/w_n)\cdot n^{-1}\sum_{k=1}^n\lambda(k)$ with $n/w_n\le1$, and $n^{-1}\sum_{k=1}^n\lambda(k)$, the Ces\`aro mean of $\lambda(1),\dots,\lambda(n)$, tends to $0$ since $\lambda(k)\to0$. We say that the normal state is \emph{linearly stable} if $Q(v)\ge0$ for all $v$, i.e.\ if $\lambda\,\Sp\mathsf K[a]\le1$. This includes the marginal case $\lambda\Sp\mathsf K[a]=1$, in which $Q$ has a non-trivial null vector.

\section{Main results}\label{sec:results}

Throughout, $\lambda$ is an interaction (Definition~\ref{def:kernel}): $\lambda(k)>0$, $\lambda(k+1)\le\lambda(k)$ and $\lambda(k)\le Ck^{-\gamma}$ for all $k\ge1$, with constants $C,\gamma>0$. Positivity and monotonicity give the coefficient signs $W_{nm}>0$ and $D_{nm}\ge0$ of~\eqref{eq:coeffs}, on which the folding and convexity arguments rest; the decay bound gives the convergence of all series and the compactness of Lemma~\ref{lem:schur}; $\lambda(0)$ never enters. $K$ is the functional~\eqref{eq:K} on admissible chains. We call $\Theta$ a \emph{minimizer} if $K(\Theta)\le K(\Theta')$ for every admissible $\Theta'$.

\begin{theorem}[Existence and uniqueness up to gauge]\label{thm:exist}
$K$ has a minimizer. There is a unique $\Theta^*\in[0,\pi/2]^{\NN_0}$ such that the minimizers are exactly $\Theta^*$ and $-\Theta^*$ (modulo $2\pi$). Either $\Theta^*=0$, or $0<\theta^*_n<\pi/2$ for every $n$.
\end{theorem}

\begin{theorem}[Linear stability decides]\label{thm:stab}
\begin{enumerate}
\item[(a)] For every admissible $\Theta$,
\begin{equation}\label{eq:KQ}
K(\Theta)=Q(\sin\Theta)+\frac12\sum_nw_n(1-\cos\theta_n)^2+\frac14\sum_{n\ne m}D_{nm}(\cos\theta_n-\cos\theta_m)^2 .
\end{equation}
\item[(b)] If the normal state is linearly stable, then $K(\Theta)>0$ for every admissible $\Theta\not\equiv0$: the normal state is the unique minimizer.
\item[(c)] If it is linearly unstable, then $\min K<0$, and the minimizers $\pm\Theta^*$ have $0<\theta^*_n<\pi/2$ for every $n$.
\end{enumerate}
\end{theorem}

The nonlinear Euler--Lagrange equations of $K$ are KAY's~\cite[eq.~(35)]{KAY2025II}, \cite[eq.~(36)]{KAY2025I}
\begin{equation}\label{eq:ELK}
w_n\sin\theta_n=\sum_{m\ge0}\Big[\lambda(n+m+1)\sin(\theta_n+\theta_m)-[m\ne n]\,\lambda(|n-m|)\sin(\theta_n-\theta_m)\Big],\quad n\in\NN_0,
\end{equation}
the sum over $m$ taken of the bracket. By the addition theorem and Lemma~\ref{lem:Dsum}, for admissible $\Theta$ they are equivalent to
\begin{equation}\label{eq:EL}
x_n\Big[w_n+2\Lambda_nc_n-\sum_{m\ne n}D_{nm}(c_n-c_m)\Big]=c_n\sum_{m\ge0}W_{nm}x_m,\qquad n\in\NN_0,
\end{equation}
in which every series converges absolutely (Section~\ref{sec:identities}). By~\cite{YuzbashyanAltshuler2022SpinChain} they are the isotropic Matsubara Eliashberg equations for even gap functions, with $\tan\theta_n=\Delta_n/\omega_n$; an admissible solution with $\theta_n\in(0,\pi/2)$ for all $n$ is a solution with $\Delta_n>0$ for all $n\ge0$.

\begin{theorem}[The positive Eliashberg solution]\label{thm:EL}
Let $\Theta$ be an admissible solution of~\eqref{eq:EL} with $\theta_n\in[0,\pi/2]$ for all $n$. Then $\Theta=0$ or $\Theta=\Theta^*$. Consequently a solution with all $\theta_n\in(0,\pi/2]$ exists if and only if the normal state is linearly unstable; it is then unique, it is the minimizer $\Theta^*$, and it has $0<\theta_n<\pi/2$ for all $n$.
\end{theorem}

\begin{theorem}[Continuity in temperature]\label{thm:cont}
Let $J\subset(0,\infty)$ be an interval and $(\lambda_T)_{T\in J}$ a family of interactions such that $T\mapsto\lambda_T(k)$ is continuous for every $k$, and such that the constants $C,\gamma$ of Definition~\ref{def:kernel} can be chosen uniformly for $T$ in each compact subset of $J$. Let $\Theta^*_T$ be the minimizer in $[0,\pi/2]^{\NN_0}$ and $\rho^*_T=\sin^2\Theta^*_T$. Then $T\mapsto\min K_T$ is continuous, and $T\mapsto\rho^*_T$ is continuous in the norm $\|\rho\|_w:=\sum_nw_n|\rho_n|$. In particular, if $T_j\to T_0$ and the normal state is linearly stable at $T_0$, then $\|\rho^*_{T_j}\|_w\to0$.
\end{theorem}

The three families of Section~\ref{sec:setting} satisfy the hypotheses of Theorem~\ref{thm:cont} on $J=(0,\infty)$: for spectra with finite second moment by dominated convergence and the bound $\lambda\langle\omega^2\rangle(2\pi T)^{-2}k^{-2}$, and for the $\gamma$-model trivially. In terms of the gap, $\|\rho^*\|_w=\sum_nw_n\Delta_n^2/(\omega_n^2+\Delta_n^2)$; it controls every $\Delta_n$.

\begin{corollary}[The thermodynamic conjectures of KAY]\label{cor:kay}
\begin{enumerate}
\item[(a)] \emph{$\gamma$-model, every $\gamma>0$.} Let $\kappa_\gamma:=\Sp\mathsf K[(j^{-\gamma})_{j\ge1}]>0$ and $T_c(\gamma):=g\kappa_\gamma^{1/\gamma}/2\pi$. For $T\ge T_c(\gamma)$ the normal state is the unique minimizer; for $T<T_c(\gamma)$ the minimizers are $\pm\Theta^*_T$ with $\Theta^*_T\ne0$; and $\|\sin^2\Theta^*_T\|_w\to0$ as $T\uparrow T_c(\gamma)$. This is Conjecture~1 of~\cite{KAY2025I}.
\item[(b)] \emph{Phonon spectra.} Let $P$ have finite second moment and $\lambda>0$, and suppose that $T\mapsto\Sp\mathsf K[a(P,T)]$ is strictly decreasing, with limits $\infty$ at $T\to0$ and $0$ at $T\to\infty$. Then the same statements hold with $T_c=T_c(\lambda,P)$ the unique solution of $\lambda\Sp\mathsf K[a(P,T_c)]=1$. The hypothesis is the theorem on uniqueness of $T_c$ at every coupling in the companion manuscript described in Section~\ref{sec:overview}, also proved in the archived preprint of Kondratev~\cite{Kondratev2026}; KAY proved uniqueness of the stability crossing, using monotonicity on a restricted temperature range, for $\lambda$ above a spectrum-dependent threshold~\cite{KAY2025II} and for Einstein phonons with $\lambda>0.7670\ldots$~\cite[Thm.~5]{KAY2025III}. In those ranges the conclusions hold without the global hypothesis, since the proof uses it only through $\{T:\lambda\Sp\mathsf K[a(P,T)]\le1\}=[T_c,\infty)$, which a unique crossing implies by the continuity of $T\mapsto\mathsf K[a(P,T)]$ (Section~\ref{sec:proofs}) and the two limits. This gives Conjecture~1 of~\cite{KAY2025II} and Conjecture~2 of~\cite{KAY2025III}.
\end{enumerate}
\end{corollary}

\begin{proof}
(a) For the $\gamma$-model, $Q_T(v)=\frac12\ip{\eta}{(I-\beta^\gamma\mathsf K[(j^{-\gamma})])\eta}$ with $\beta=g/2\pi T$, by~\eqref{eq:QK}. $\mathsf K[(j^{-\gamma})]$ is compact and self-adjoint with $(0,0)$ entry $1$ (since $A_0=0$), so $\kappa_\gamma\ge1$ is its largest eigenvalue, and $Q_T\ge0$ if and only if $\beta^\gamma\kappa_\gamma\le1$, i.e.\ $T\ge T_c(\gamma)$. Theorems~\ref{thm:exist}--\ref{thm:cont} give the rest. (b) By~\eqref{eq:QK} the normal state is linearly stable if and only if $\lambda\Sp\mathsf K[a(P,T)]\le1$, which under the hypothesis holds exactly for $T\ge T_c$. KAY's admissible spectra have finite second moment, as noted in Section~\ref{sec:setting}.
\end{proof}

Without any monotonicity hypothesis, Theorems~\ref{thm:exist}--\ref{thm:cont} say: at every temperature the minimizer is unique up to gauge; the set of temperatures at which it is the normal state is the set where $\lambda\Sp\mathsf K[a(P,T)]\le1$; and the minimizer is continuous in $T$, so that it vanishes continuously at every boundary point of that set.

\begin{remark}[Finite truncations]\label{rem:finite}
Theorems~\ref{thm:exist}--\ref{thm:cont} hold for the chain truncated to $N$ spins, with the same proofs except that existence follows from continuity on the compact cube $[0,1]^N$ and no compactness or Fatou argument is needed. There are two natural truncations. Put $t^{(N)}_n:=\sum_{m\ge N}D_{nm}$ and $\Lambda^{(N)}_n:=\frac12\big(\sum_{m<N,m\ne n}\lambda(|n-m|)-\sum_{m<N}\lambda(n+m+1)\big)=\Lambda_n-\frac12t^{(N)}_n$. Deleting the spins $n\ge N$ replaces $\Lambda_n$ by $\Lambda^{(N)}_n$ in all identities. Freezing them in the normal state replaces in addition $w_n$ by $w_n+t^{(N)}_n$ in every term carrying the weight $w_n$; then $Q$ becomes the upper-left $N\times N$ block of~\eqref{eq:Q}. The coefficient $\Lambda^{(N)}_n$ can be negative, which affects only the non-negativity statements used for coercivity: it multiplies a term linear in $\rho$ and plays no role in convexity. The Lean formalization treats these finite chains with arbitrary positive weights and arbitrary linear coefficients.
\end{remark}

\section{Exact identities and the gauge reduction}\label{sec:identities}

\begin{lemma}[Rearrangement]\label{lem:identity}
For admissible $\Theta$,
\begin{equation}\label{eq:ident}
K(\Theta)=\sum_nw_n(1-c_n)+\sum_n\Lambda_n\rho_n+\frac14\sum_{n\ne m}D_{nm}(c_n-c_m)^2-\frac12\sum_{n,m}W_{nm}x_nx_m,
\end{equation}
where all four series converge absolutely and the first three have non-negative terms.
\end{lemma}

\begin{proof}
By~\eqref{eq:pair} and $1-c_nc_m=\frac12(\rho_n+\rho_m)+\frac12(c_n-c_m)^2$, the off-diagonal part of $K$ in grouped form is
\[
\frac12\sum_{n\ne m}\Big[\tfrac12D_{nm}(\rho_n+\rho_m)+\tfrac12D_{nm}(c_n-c_m)^2-W_{nm}x_nx_m\Big].
\]
Each of the three pieces converges absolutely (Lemma~\ref{lem:welldef}), so it may be summed separately. By symmetry and Tonelli, $\frac14\sum_{n\ne m}D_{nm}(\rho_n+\rho_m)=\frac12\sum_n\rho_n\sum_{m\ne n}D_{nm}=\sum_n\rho_n\big(\Lambda_n+\frac12\lambda(2n+1)\big)$ by Lemma~\ref{lem:Dsum}. The diagonal part of $K$ is $-\frac12\sum_n\lambda(2n+1)(1-\cos2\theta_n)=-\sum_n\lambda(2n+1)\rho_n$. Adding, the coefficient of $\rho_n$ is $\Lambda_n-\frac12\lambda(2n+1)=\Lambda_n-\frac12W_{nn}$, and $-\frac12W_{nn}\rho_n=-\frac12W_{nn}x_n^2$ is the diagonal of the last sum in~\eqref{eq:ident}.
\end{proof}

\begin{proof}[Proof of Theorem~\ref{thm:stab}(a)]
Since $1-c_n-\frac12x_n^2=1-c_n-\frac12(1-c_n^2)=\frac12(1-c_n)^2$, Lemma~\ref{lem:identity} gives~\eqref{eq:KQ}, with $Q(\sin\Theta)$ finite.
\end{proof}

The Euler--Lagrange equations~\eqref{eq:EL} follow from~\eqref{eq:ELK} in the same way: with $\sin(\theta_n\pm\theta_m)=x_nc_m\pm c_nx_m$, the bracket in~\eqref{eq:ELK} equals $c_nW_{nm}x_m-x_nD_{nm}c_m$ for $m\ne n$ and $\lambda(2n+1)(x_nc_n+c_nx_n)$ for $m=n$. Writing $-x_nD_{nm}c_m=x_nD_{nm}(c_n-c_m)-x_nD_{nm}c_n$ and using Lemma~\ref{lem:Dsum} gives~\eqref{eq:EL}; the series $\sum_mW_{nm}|x_m|$ and $\sum_{m\ne n}D_{nm}|c_n-c_m|$ converge, the first because $W_{nm}\le2\bar\lambda(|n-m|)$ and, by Cauchy--Schwarz, $\sum_m\bar\lambda(|n-m|)|x_m|\le\|\xi\|\big(\sum_m\bar\lambda(|n-m|)^2/w_m\big)^{1/2}<\infty$, the second by Lemma~\ref{lem:Dsum}.

\begin{lemma}[Gauge reduction]\label{lem:fold}
For admissible $\Theta$ let $\hat\theta_n:=\arcsin|\sin\theta_n|\in[0,\pi/2]$, so that $\sin\hat\theta_n=|x_n|$ and $\cos\hat\theta_n=|c_n|$. Then $\hat\Theta$ is admissible and $K(\hat\Theta)\le K(\Theta)$. Equality holds if and only if $\Theta=\hat\Theta$ or $\Theta=-\hat\Theta$ modulo $2\pi$.
\end{lemma}

\begin{proof}
$1-|c_n|\le1-c_n$, so $\hat\Theta$ is admissible. Compare~\eqref{eq:ident} term by term: $w_n(1-|c_n|)\le w_n(1-c_n)$, the $\rho_n$ are unchanged, $(|c_n|-|c_m|)^2\le(c_n-c_m)^2$ with $D_{nm}\ge0$, and $-W_{nm}|x_n||x_m|\le-W_{nm}x_nx_m$ with $W_{nm}>0$. If equality holds, then $c_n\ge0$ for all $n$ (else the first comparison is strict at $n$, as $w_n>0$), and $x_nx_m\ge0$ for all $n,m$ (else the last is strict). So the non-zero $x_n$ have a common sign, and $\theta_n\in[0,\pi/2]$ for all $n$, or $\theta_n\in[-\pi/2,0]$ for all $n$, modulo $2\pi$. The converse is clear since $K$ is even.
\end{proof}

That a minimizer can be taken in $[0,\pi/2]^{\mathbb N_0}$ was shown by Yuzbashyan and Altshuler~\cite{YuzbashyanAltshuler2022SpinChain} (see~\cite[Sect.~4]{KAY2025II}); we include the short proof because its equality case is used below.

\section{Convexity}\label{sec:convex}

Let $\Rset:=\{\rho\in[0,1]^{\NN_0}:\sum_nw_n\rho_n<\infty\}$, a convex set. The map $\Theta\mapsto\sin^2\Theta$ is a bijection from admissible $\Theta\in[0,\pi/2]^{\NN_0}$ onto $\Rset$, with inverse $\rho\mapsto(\arcsin\sqrt{\rho_n})_n$, because $\frac12\rho\le1-\sqrt{1-\rho}\le\rho$ on $[0,1]$. Put $g(r):=1-\sqrt{1-r}$ and, for $\rho\in\Rset$, $F(\rho):=K\big((\arcsin\sqrt{\rho_n})_n\big)$. By Lemma~\ref{lem:identity},
\begin{multline}\label{eq:F}
F(\rho)=\sum_nw_ng(\rho_n)+\sum_n\Lambda_n\rho_n+\frac14\sum_{n\ne m}D_{nm}\big(\sqrt{1-\rho_n}-\sqrt{1-\rho_m}\big)^2\\
-\frac12\sum_{n,m}W_{nm}\sqrt{\rho_n\rho_m}.
\end{multline}

\begin{lemma}\label{lem:convfns}
\begin{enumerate}
\item[(a)] $g$ is strictly convex on $[0,1]$, increasing, and $g'(r)=\frac12(1-r)^{-1/2}$ on $[0,1)$.
\item[(b)] $(a,b)\mapsto\sqrt{ab}$ is concave on $[0,\infty)^2$.
\item[(c)] $(a,b)\mapsto(\sqrt{1-a}-\sqrt{1-b})^2$ is convex on $[0,1]^2$.
\end{enumerate}
\end{lemma}

\begin{proof}
(a) $\sqrt{1-r}$ is strictly concave. (b) For $t\in[0,1]$, by Cauchy--Schwarz, $t\sqrt{ab}+(1-t)\sqrt{a'b'}=\sqrt{ta}\sqrt{tb}+\sqrt{(1-t)a'}\sqrt{(1-t)b'}\le\sqrt{ta+(1-t)a'}\sqrt{tb+(1-t)b'}$. (c) The function equals $(1-a)+(1-b)-2\sqrt{(1-a)(1-b)}$, an affine function minus a concave function of the affine arguments $(1-a,1-b)$, by (b).
\end{proof}

\begin{proposition}[Hidden convexity]\label{prop:convex}
$F$ is strictly convex on $\Rset$.
\end{proposition}

\begin{proof}
Let $\rho,\rho'\in\Rset$, $t\in(0,1)$ and $\rho^t=t\rho+(1-t)\rho'$. Each term of each series in~\eqref{eq:F} is a convex function of $\rho$ by Lemma~\ref{lem:convfns}, with $D_{nm}\ge0$ and $W_{nm}\ge0$; and each series converges absolutely at $\rho,\rho'$ and $\rho^t$ (Lemma~\ref{lem:welldef}). Summing the termwise inequalities gives $F(\rho^t)\le tF(\rho)+(1-t)F(\rho')$. If $\rho\ne\rho'$, then $\rho_n\ne\rho'_n$ for some $n$, and the inequality for the term $w_ng(\rho_n)$ is strict by Lemma~\ref{lem:convfns}(a).
\end{proof}

For $\rho\in\Rset$ write $\xi(\rho)_n:=\sqrt{w_n\rho_n}$, so that $\|\xi(\rho)\|^2=\sum_nw_n\rho_n$ and $\frac12\sum_{n,m}W_{nm}\sqrt{\rho_n\rho_m}=\frac12\ip{\xi(\rho)}{M\xi(\rho)}$.

\section{Existence}\label{sec:exist}

\begin{lemma}[Coercivity]\label{lem:coerc}
Choose $N$ with $\tau_N\le\frac12$ (Lemma~\ref{lem:schur}). Then for every $\rho\in\Rset$
\[
F(\rho)\ge\tfrac14\|\xi(\rho)\|^2-\tfrac12\|M\|N^2 .
\]
\end{lemma}

\begin{proof}
The second and third series in~\eqref{eq:F} are non-negative and $w_ng(\rho_n)\ge\frac12w_n\rho_n$, so $F(\rho)\ge\frac12\|\xi\|^2-\frac12\ip{\xi}{M\xi}$. Now $\ip{\xi}{M\xi}\le\tau_N\|\xi\|^2+\|M\|\,\|P_N\xi\|^2$ and $\|P_N\xi\|^2=\sum_{n<N}w_n\rho_n\le N^2$.
\end{proof}

\begin{lemma}[Lower semicontinuity]\label{lem:lsc}
Let $\rho^j\in\Rset$ with $\sup_j\|\xi(\rho^j)\|<\infty$ and $\rho^j_n\to\rho_n$ for every $n$. Then $\rho\in\Rset$, $\xi(\rho^j)\rightharpoonup\xi(\rho)$ weakly in $\ell^2$, and $F(\rho)\le\liminf_jF(\rho^j)$. More precisely, the first three series in~\eqref{eq:F} are lower semicontinuous along the sequence and the fourth converges.
\end{lemma}

\begin{proof}
$\rho_n\in[0,1]$, and $\sum_nw_n\rho_n\le\liminf_j\sum_nw_n\rho^j_n<\infty$ by Fatou, so $\rho\in\Rset$. A bounded sequence in $\ell^2$ that converges coordinatewise converges weakly to the coordinatewise limit. The first three series have non-negative terms that converge termwise, so Fatou's lemma applies. Since $M$ is compact, $M\xi(\rho^j)\to M\xi(\rho)$ in norm, and $\ip{\xi(\rho^j)}{M\xi(\rho^j)}\to\ip{\xi(\rho)}{M\xi(\rho)}$.
\end{proof}

\begin{proof}[Proof of Theorem~\ref{thm:exist}]
By Lemma~\ref{lem:coerc}, $\inf_\Rset F>-\infty$. Let $\rho^j$ be a minimizing sequence. By Lemma~\ref{lem:coerc} again, $\|\xi(\rho^j)\|$ is bounded. By a diagonal argument a subsequence converges coordinatewise to some $\rho^*$, and by Lemma~\ref{lem:lsc}, $\rho^*\in\Rset$ and $F(\rho^*)\le\inf F$. So $\rho^*$ minimizes $F$, and it is the unique minimizer by Proposition~\ref{prop:convex}. Let $\Theta^*=(\arcsin\sqrt{\rho^*_n})_n$. For any admissible $\Theta$, Lemma~\ref{lem:fold} gives $K(\Theta)\ge K(\hat\Theta)=F(\sin^2\hat\Theta)\ge F(\rho^*)=K(\Theta^*)$. So $\Theta^*$ is a minimizer. If $\Theta$ is a minimizer, equality holds throughout, so $\sin^2\hat\Theta=\rho^*$, i.e.\ $\hat\Theta=\Theta^*$, and $\Theta=\pm\Theta^*$ by the equality case of Lemma~\ref{lem:fold}. The last assertion is Lemma~\ref{lem:interior} below.
\end{proof}

\begin{lemma}[Interior]\label{lem:interior}
If $\rho^*\ne0$ minimizes $F$ on $\Rset$, then $0<\rho^*_n<1$ for every $n$.
\end{lemma}

\begin{proof}
First, $\sigma_n:=\sum_mW_{nm}\sqrt{\rho^*_m}<\infty$ for every $n$, by the Cauchy--Schwarz estimate after~\eqref{eq:EL}.

Suppose $\rho^*_n=0$ for some $n$. Since $\rho^*\ne0$ and $W>0$, $\sigma_n>0$. Let $0<s<1$ and $\rho=\rho^*+se_n\in\Rset$. Compare~\eqref{eq:F} at $\rho$ and $\rho^*$. The first two series increase by $w_ng(s)+\Lambda_ns\le(w_n+\Lambda_n)s$. In the third, only the terms with $n$ as one index change, by $\frac12\sum_{m\ne n}D_{nm}\big[(\sqrt{1-s}-c_m)^2-(1-c_m)^2\big]$ with $c_m=\sqrt{1-\rho^*_m}$; since $(a-c)^2-(1-c)^2=(a-1)(a+1-2c)$ and $|a-1|\le s$ for $a=\sqrt{1-s}$, this is at most $s\sum_{m\ne n}D_{nm}\le w_n\lambda(1)s$. The fourth changes by $-\sqrt s\sum_{m\ne n}W_{nm}\sqrt{\rho^*_m}-\frac12W_{nn}s=-\sigma_n\sqrt s-\frac12W_{nn}s$. Hence $F(\rho)-F(\rho^*)\le C_ns-\sigma_n\sqrt s<0$ for small $s$, a contradiction.

Suppose $\rho^*_n=1$. Let $\rho=\rho^*-se_n$. The first series decreases by $w_n\sqrt s$, the second does not increase. The third changes by $\frac12\sum_{m\ne n}D_{nm}\big[(\sqrt s-c_m)^2-c_m^2\big]\le\frac12s\sum_{m\ne n}D_{nm}$. The fourth changes by $\sum_{m\ne n}W_{nm}\sqrt{\rho^*_m}\,(1-\sqrt{1-s})+\frac12W_{nn}s\le(\sigma_n+W_{nn})s$. Hence $F(\rho)-F(\rho^*)\le C'_ns-w_n\sqrt s<0$ for small $s$, a contradiction.
\end{proof}

\section{Proofs of Theorems~\ref{thm:stab}--\ref{thm:cont}}\label{sec:proofs}

\begin{proof}[Proof of Theorem~\ref{thm:stab}(b),(c)]
(b) If $Q\ge0$, then by~\eqref{eq:KQ} $K(\Theta)\ge\frac12\sum_nw_n(1-\cos\theta_n)^2$, which vanishes only if $\cos\theta_n=1$ for all $n$.

(c) Suppose $Q(v)<0$ for some $v$ with $\sum w_nv_n^2<\infty$. Since $W>0$, $Q(|v|)\le Q(v)$, so we may take $v\ge0$. Truncations $v^{(N)}=P_Nv$ satisfy $Q(v^{(N)})\to Q(v)$ (the first two series converge monotonically, and $\eta^{(N)}\to\eta$ in $\ell^2$ with $M$ bounded), so we may take $v$ finitely supported. For $0<t\le1/\max_nv_n$ let $\theta_n=\arcsin(tv_n)$, so $x_n=tv_n$ and $1-c_n=x_n^2/(1+c_n)\le t^2v_n^2$. By~\eqref{eq:KQ}, using $(c_n-c_m)^2\le\big((1-c_n)+(1-c_m)\big)^2\le t^4(v_n^2+v_m^2)^2\le2t^4(v_n^4+v_m^4)$ and Lemma~\ref{lem:Dsum},
\[
K(\Theta)\le t^2Q(v)+\tfrac12t^4\sum_nw_nv_n^4+t^4\lambda(1)\sum_nw_nv_n^4,
\]
which is negative for small $t$. So $\min K<0$, the minimizer is not the normal state, and Theorem~\ref{thm:exist} gives $0<\theta^*_n<\pi/2$.
\end{proof}

\subsection{Stationary points}

Let $U\subset\Rset$ be the set of $\rho\in\Rset$ with $0<\rho_n<1$ for all $n$.

\begin{lemma}[Partial derivatives]\label{lem:partial}
Let $\rho^0\in U$, let $S\subset\NN_0$ be finite, and let $V$ be the set of $\rho\in\Rset$ that agree with $\rho^0$ outside $S$ and have $\rho_n\in(0,1)$ for $n\in S$. On $V$, regarded as an open subset of $\RR^S$, $F$ is continuously differentiable, with
\begin{equation}\label{eq:partial}
\partial_nF(\rho)=\frac{w_n}{2c_n}+\Lambda_n-\frac1{2c_n}\sum_{m\ne n}D_{nm}(c_n-c_m)-\frac1{2x_n}\sum_mW_{nm}x_m,\qquad n\in S,
\end{equation}
where $x_m=\sqrt{\rho_m}$ and $c_m=\sqrt{1-\rho_m}$. Moreover $2x_nc_n\,\partial_nF(\rho)$ is the left side minus the right side of~\eqref{eq:EL}.
\end{lemma}

\begin{proof}
Only finitely many terms of the first two series in~\eqref{eq:F} depend on $(\rho_n)_{n\in S}$. In the third and fourth, a term depends on $\rho_n$, $n\in S$, only if $n$ is one of its two indices. Differentiate termwise: for fixed $n\in S$ the series of derivatives are $-\frac1{2c_n}\sum_{m\ne n}D_{nm}(c_n-c_m)$ (each unordered pair $\{n,m\}$ appears twice in~\eqref{eq:F}) and $-\frac1{2x_n}\sum_{m\ne n}W_{nm}x_m-\frac12W_{nn}$. On any compact subset of $V$, $c_n$ and $x_n$ for $n\in S$ are bounded away from $0$, $|c_n-c_m|\le1$, and $x_m=\sqrt{\rho^0_m}$ for $m\notin S$; so these series converge uniformly, with majorants $\sum_mD_{nm}$ (Lemma~\ref{lem:Dsum}) and $\sum_mW_{nm}\sqrt{\rho^0_m}+\sum_{m\in S}W_{nm}$ (Lemma~\ref{lem:interior}, first paragraph, which uses only $\rho^0\in\Rset$). Termwise differentiation of a series of $C^1$ functions with locally uniformly convergent derivative series is legitimate, and the limit is continuous. Multiplying by $2x_nc_n$ gives the last statement.
\end{proof}

\begin{proof}[Proof of Theorem~\ref{thm:EL}]
Let $\Theta$ be an admissible solution of~\eqref{eq:EL} in $[0,\pi/2]^{\NN_0}$, $\Theta\ne0$. If $x_n=0$ for some $n$, then $c_n=1$ and~\eqref{eq:EL} gives $\sum_mW_{nm}x_m=0$, so $x\equiv0$ since $W>0$ and $x\ge0$, a contradiction. If $c_n=0$, then $x_n=1$ and the left side of~\eqref{eq:EL} is $w_n+\sum_{m\ne n}D_{nm}c_m>0$ while the right side is $0$. Hence $\rho_s:=\sin^2\Theta\in U$, and $\partial_nF(\rho_s)=0$ for all $n$ by Lemma~\ref{lem:partial}.

\emph{Step 1: finitely supported perturbations.} Let $\rho\in\Rset$ agree with $\rho_s$ outside a finite set $S$. The function $\varphi(t)=F(\rho_s+t(\rho-\rho_s))$ on $[0,1]$ is convex by Proposition~\ref{prop:convex}. For small $t\ge0$ the point $\rho_s+t(\rho-\rho_s)$ lies in the set $V$ of Lemma~\ref{lem:partial} built on $\rho_s$, so $\varphi$ is differentiable from the right at $0$ with $\varphi'(0^+)=\sum_{n\in S}(\rho_n-\rho_{s,n})\partial_nF(\rho_s)=0$. A convex function on $[0,1]$ with $\varphi'(0^+)=0$ satisfies $\varphi(1)\ge\varphi(0)$, i.e.\ $F(\rho)\ge F(\rho_s)$.

\emph{Step 2: approximation.} Let $\rho\in\Rset$ be arbitrary, and let $\rho^N$ agree with $\rho$ for $n<N$ and with $\rho_s$ for $n\ge N$. Then $\rho^N\in\Rset$, and by Step~1, $F(\rho^N)\ge F(\rho_s)$. We show $F(\rho^N)\to F(\rho)$. The first two series in~\eqref{eq:F} at $\rho^N$ are $\sum_{n<N}(\cdots)(\rho)+\sum_{n\ge N}(\cdots)(\rho_s)$, which converges to their value at $\rho$. In the third, the terms converge for each fixed $(n,m)$ once $N>\max(n,m)$, and they are dominated by $2D_{nm}\big((1-c_n)+(1-c_{s,n})+(1-c_m)+(1-c_{s,m})\big)$, because $(c^N_n-c^N_m)^2\le2(1-c^N_n)+2(1-c^N_m)$ and $1-c^N_n\le(1-c_n)+(1-c_{s,n})$; the majorant is summable by Lemma~\ref{lem:Dsum}. In the fourth, $\xi(\rho^N)-\xi(\rho)$ vanishes for $n<N$ and is bounded by $\xi(\rho)_n+\xi(\rho_s)_n$ for $n\ge N$, so $\xi(\rho^N)\to\xi(\rho)$ in $\ell^2$, and $M$ is bounded. Hence $F(\rho)\ge F(\rho_s)$.

So $\rho_s$ minimizes $F$ on $\Rset$; by Proposition~\ref{prop:convex} and the proof of Theorem~\ref{thm:exist}, $\rho_s=\rho^*$ and $\Theta=\Theta^*$.

It remains to show that a solution with all $\theta_n\in(0,\pi/2]$ exists if and only if the normal state is linearly unstable. Such a solution is not $0$, so by the above it equals $\Theta^*\ne0$, and then the normal state is unstable by Theorem~\ref{thm:stab}(b). Conversely, if the normal state is unstable, then $\Theta^*\ne0$ and $\rho^*\in U$ by Theorem~\ref{thm:stab}(c). Since $\rho^*$ minimizes $F$ on the set $V$ of Lemma~\ref{lem:partial} built on $\rho^0=\rho^*$, for any finite $S$, all partial derivatives vanish there, and by the last statement of Lemma~\ref{lem:partial}, $\Theta^*$ solves~\eqref{eq:EL}.
\end{proof}

\begin{proof}[Proof of Theorem~\ref{thm:cont}]
Fix a compact interval $J_0\subset J$ and constants $C,\gamma$ valid on $J_0$. The bounds of Lemmas~\ref{lem:schur} and~\ref{lem:coerc} are then uniform on $J_0$: with $\tau_N$ and $\|M_T\|\le2\sigma_0$ depending only on $C,\gamma$, $F_T(\rho)\ge\frac14\|\xi(\rho)\|^2-B$ for $T\in J_0$, with $B$ independent of $T$.

\emph{Operator continuity.} For $T,T'\in J_0$, $\|M_T-M_{T'}\|\le2\tau_N+\|P_N(M_T-M_{T'})P_N\|$, and the finite matrix depends continuously on $T$; so $T\mapsto M_T$ is norm continuous.

\emph{Pointwise continuity of $F_T$.} Fix $\rho\in\Rset$. The first series in~\eqref{eq:F} does not depend on $T$. The second converges by dominated convergence, with majorant $nC\rho_n$. In the third, the terms $f_{T,nm}=\frac14D_{T,nm}(c_n-c_m)^2$ satisfy $0\le f_{T,nm}\le h_{T,nm}:=\frac12D_{T,nm}\big((1-c_n)+(1-c_m)\big)$, because $(c_n-c_m)^2\le2(1-c_n)^2+2(1-c_m)^2$ and $0\le1-c_n\le1$. By Lemma~\ref{lem:Dsum}, $\sum_{n\ne m}h_{T,nm}=\sum_n(1-c_n)\big(2\Lambda_{T,n}+\lambda_T(2n+1)\big)$, which is continuous in $T$ by dominated convergence with majorant $(1-c_n)w_nC$. Fatou's lemma applied to $f_T\ge0$ and to $h_T-f_T\ge0$, both of which converge termwise as $T\to T_0$, gives $\liminf\sum f_T\ge\sum f_{T_0}$ and $\liminf\sum(h_T-f_T)\ge\sum(h_{T_0}-f_{T_0})$; with $\sum h_T\to\sum h_{T_0}<\infty$ the second gives $\limsup\sum f_T\le\sum f_{T_0}$. The fourth series is continuous by operator continuity. So $T\mapsto F_T(\rho)$ is continuous.

\emph{Convergence of minimizers.} Let $T_j\to T_0$ in $J_0$ and $\rho^j=\rho^*_{T_j}$. Since $F_{T_j}(\rho^j)\le F_{T_j}(0)=0$, $\|\xi(\rho^j)\|$ is bounded. Take any subsequence; a further subsequence converges coordinatewise to some $\bar\rho\in\Rset$, and then $\xi^j:=\xi(\rho^j)\rightharpoonup\bar\xi:=\xi(\bar\rho)$ weakly in $\ell^2$. The first three series in~\eqref{eq:F} have non-negative terms that converge termwise (the coefficients $\Lambda_{T,n}$, $D_{T,nm}$ converge termwise), so by Fatou's lemma their $\liminf$ along the sequence is at least their value at $(T_0,\bar\rho)$. The fourth series enters with a negative sign, and it converges: $\ip{\xi^j}{M_{T_0}\xi^j}\to\ip{\bar\xi}{M_{T_0}\bar\xi}$ by compactness of $M_{T_0}$, and $|\ip{\xi^j}{(M_{T_j}-M_{T_0})\xi^j}|\le\|M_{T_j}-M_{T_0}\|\,\|\xi^j\|^2\to0$ by norm continuity. Together with the minimality of $\rho^j$ and the pointwise continuity of $F_T$ at $\rho^*_{T_0}$ this gives
\[
F_{T_0}(\bar\rho)\le\liminf_jF_{T_j}(\rho^j)\le\liminf_jF_{T_j}(\rho^*_{T_0})=F_{T_0}(\rho^*_{T_0}).
\]
So $\bar\rho=\rho^*_{T_0}$ by uniqueness, and the whole sequence converges coordinatewise to $\rho^*_{T_0}$. The same chain of inequalities, with $\limsup_jF_{T_j}(\rho^j)\le\lim_jF_{T_j}(\rho^*_{T_0})$, shows $\min F_{T_j}\to\min F_{T_0}$.

\emph{Norm convergence.} Write $F_T=G_1+G_{2,T}-G_{3,T}$, with $G_1(\rho)=\sum_nw_ng(\rho_n)$, $G_{2,T}$ the second and third series and $G_{3,T}$ the fourth. Along the sequence, $F_{T_j}(\rho^j)\to F_{T_0}(\rho^*_{T_0})$ and $G_{3,T_j}(\rho^j)\to G_{3,T_0}(\rho^*_{T_0})$, as just shown, so $G_1(\rho^j)+G_{2,T_j}(\rho^j)$ converges to $G_1(\rho^*_{T_0})+G_{2,T_0}(\rho^*_{T_0})$. By Fatou's lemma along the sequence (non-negative terms, termwise convergent), $\liminf_jG_1(\rho^j)\ge G_1(\rho^*_{T_0})$ and $\liminf_jG_{2,T_j}(\rho^j)\ge G_{2,T_0}(\rho^*_{T_0})$. Since the sum converges, both $\liminf$ are limits, and $G_1(\rho^j)\to G_1(\rho^*_{T_0})$. The non-negative sequences $(w_ng(\rho^j_n))_n$ converge termwise and in sum, so by Scheff\'e's lemma they converge in $\ell^1$. Since $|r-r'|\le2|g(r)-g(r')|$ on $[0,1]$ (as $g'\ge\frac12$), $\|\rho^j-\rho^*_{T_0}\|_w\to0$. If the normal state is linearly stable at $T_0$, then $\rho^*_{T_0}=0$ by Theorem~\ref{thm:stab}(b).
\end{proof}

\section{Verification}\label{sec:verify}

\subsection{Lean formalization}\label{sec:lean}

The finite-dimensional core of the argument is formalized in Lean~4~\cite{Lean4} with Mathlib~\cite{Mathlib2020}; the source is in the supplementary material (directory \texttt{lean/}, namespace \texttt{KayThermo}). The formal setting is the chain of Remark~\ref{rem:finite} with $N$ spins, an arbitrary weight vector $w\in(0,\infty)^N$ and an arbitrary sequence $\lambda:\NN_0\to\RR$ with $\lambda(k)>0$ for $k\ge1$ and $\lambda$ non-increasing on $k\ge1$; no decay or summability is needed in finite dimension. Since the weights are arbitrary, this covers both truncations of Remark~\ref{rem:finite}: the frozen one is the deleted one with $w_n$ replaced by $w_n+t^{(N)}_n$. The functional $K$ is defined in Lean exactly as the finite form of~\eqref{eq:K}, and $\Lambda_n$, $D_{nm}$, $W_{nm}$, $Q$ and the density functional $F$ by their finite formulas. The following statements are proved, without \texttt{sorry} or added axioms.
\begin{enumerate}
\item \texttt{K\_\allowbreak{}eq\_\allowbreak{}identity}: Lemma~\ref{lem:identity}, the rearrangement identity~\eqref{eq:ident}, for all real angles and with no sign hypotheses.
\item \texttt{K\_\allowbreak{}eq\_\allowbreak{}Q\_\allowbreak{}plus}: Theorem~\ref{thm:stab}(a), the decomposition~\eqref{eq:KQ}, for all real angles.
\item \texttt{normal\_\allowbreak{}unique\_\allowbreak{}min}, \texttt{unstable\_\allowbreak{}neg} and \texttt{normal\_\allowbreak{}global\_\allowbreak{}min\_\allowbreak{}iff\_\allowbreak{}Q\_\allowbreak{}nonneg}: Theorem~\ref{thm:stab} in finite dimension, that is, if $Q\ge0$ then $K\ge0$ with equality only at the normal state (marginal case included); if $Q(v)<0$ for some $v$ then $K$ takes a negative value; hence the normal state is a global minimizer if and only if $Q\ge0$.
\item \texttt{K\_\allowbreak{}fold\_\allowbreak{}le}: Lemma~\ref{lem:fold}: folding does not increase $K$, and equality implies $c_n\ge0$ for all $n$ and $x_nx_m\ge0$ for all $n,m$ (the converse direction of the lemma is not stated in Lean).
\item \texttt{K\_\allowbreak{}fold\_\allowbreak{}eq\_\allowbreak{}G}, \texttt{G\_\allowbreak{}eq\_\allowbreak{}K\_\allowbreak{}fold}: $K=F(\sin^2\Theta)$ on the first quadrant and $F(\rho)=K(\arcsin\sqrt\rho)$ on $[0,1]^N$.
\item \texttt{Gen\_\allowbreak{}strictConvexOn}, \texttt{G\_\allowbreak{}strictConvexOn}: Proposition~\ref{prop:convex}. The general form is the strict convexity on $[0,1]^N$ of the functional of~\eqref{eq:F} with an arbitrary real $L_n$ in place of $\Lambda_n$, any $w_n>0$, any $D_{nm}\ge0$ for $n\ne m$ and any $W_{nm}\ge0$.
\item \texttt{exists\_\allowbreak{}min}, \texttt{min\_\allowbreak{}unique\_\allowbreak{}up\_\allowbreak{}to\_\allowbreak{}gauge}: Theorem~\ref{thm:exist} in finite dimension: a minimizer over all real angle vectors exists, and any two minimizers $\Theta,\Theta'$ satisfy $\cos\theta_n=\cos\theta'_n$ for all $n$ and either $\sin\theta_n=\sin\theta'_n$ for all $n$ or $\sin\theta_n=-\sin\theta'_n$ for all $n$.
\item \texttt{G\_\allowbreak{}stationary\_\allowbreak{}is\_\allowbreak{}min}, \texttt{G\_\allowbreak{}stationary\_\allowbreak{}unique}: a point of $[0,1]^N$ at which $F$ has zero Fréchet derivative is a global minimizer of $F$ on $[0,1]^N$, and there is at most one such point. This is the convex-analytic core of Theorem~\ref{thm:EL}.
\end{enumerate}
The build (\texttt{lake build}) completes, and for each theorem listed \texttt{\#print axioms} reports only \texttt{propext}, \texttt{Classical.choice} and \texttt{Quot.sound}. Not formalized are the infinite-chain analysis (Lemma~\ref{lem:schur}, Lemmas~\ref{lem:coerc}--\ref{lem:interior}, absolute convergence and the density argument of Section~\ref{sec:proofs}), the translation between the Eliashberg equations~\eqref{eq:ELK} and the vanishing of the derivative of $F$ (Lemma~\ref{lem:partial} and the interior property), the continuity Theorem~\ref{thm:cont}, and the spectral input of Corollary~\ref{cor:kay}.

\subsection{Numerical checks}\label{sec:numerics}

The numerical checks are finite-truncation observations and are not used in any proof. We took seven interactions: Einstein phonons with $\lambda\in\{0.5,1,3\}$ ($\Omega=1$); a two-peak spectrum $P=\frac12(\delta_1+\delta_3)$ with $\lambda=1$; and the $\gamma$-model with $\gamma\in\{0.5,1,2\}$ ($g=1$). The chain was truncated to $N$ spins with the exterior spins frozen in the normal state (Remark~\ref{rem:finite}). For an interaction $\lambda(k)=\lambda a_k$ let $\mathsf K_N$ be the $N\times N$ principal block of $\lambda\mathsf K[a]$, that is, the block of~\eqref{eq:QK} with $\lambda a_k$ read as $\lambda(k)$; for the $\gamma$-model we took the scalar $\lambda=1$ and $a_k=\lambda(k)$. With $\mathcal W_N=\operatorname{diag}(w_0,\dots,w_{N-1})$, the angle Hessian at $\Theta=0$ of the truncated chain is then $\mathcal W_N^{1/2}(I-\mathsf K_N)\mathcal W_N^{1/2}$. We computed the stability temperature $T_c^{(N)}$, the temperature at which the largest eigenvalue of $\mathsf K_N$ reaches $1$, by Brent's method in $\log T$, for $N\in\{64,128,256,400\}$. For the phonon cases and $\gamma=2$ the values at $N=256$ and $N=400$ agree to at least seven significant digits (for Einstein phonons with $\lambda=1$, $T_c^{(400)}/\Omega=0.1146195121$); for $\gamma=1$ the relative difference is $7.3\times10^{-6}$, and for $\gamma=0.5$ the two still differ in the third digit, as expected for a non-summable interaction.

At $N=64$ and $T/T_c^{(64)}\in\{0.5,0.8,1.1,1.4\}$ we minimized $K$ from 30 random starts (15 in the variables $\rho$, 15 in the angles), folded each result to the first quadrant (Lemma~\ref{lem:fold}) and polished it by a second bounded minimization followed by a Powell hybrid root solve of the gradient, with the analytic angle Hessian as Jacobian; where a run stalled at the unstable normal state below $T_c$ it was restarted in the direction $(|u_n|/\sqrt{w_n})_n$, with $u$ a leading eigenvector of $\mathsf K_N$, before the root solve. After folding and polishing, all runs agreed with the best candidate within $9.1\times10^{-13}$ in $\rho$ in all 28 settings, and gave the normal state for $T>T_c^{(64)}$. (Before folding, the raw optimizer runs did not all agree: the angle landscape is not convex.) We also solved the Eliashberg equations for $(Z_n,\Delta_n)$ by the standard fixed-point iteration with one-half damping from three starts; the solutions agreed with the minimizer to $5\times10^{-11}$ in the angles. A Powell hybrid root solve from eight random starts uniform in $[-\pi,\pi]^N$, with a finite-difference Jacobian, found further solutions of~\eqref{eq:ELK}, many of them sign-changing, all with energies far above the minimum; none had lower energy than the minimizer. In every case the minimizer $\rho^*_n$ was non-increasing in $n$, and as $N$ increased from $64$ to $400$ at $T=0.7\,T_c^{(N)}$ the minimum energy decreased monotonically in $N$.

\begin{figure}[t]
\centering
\includegraphics[width=\textwidth]{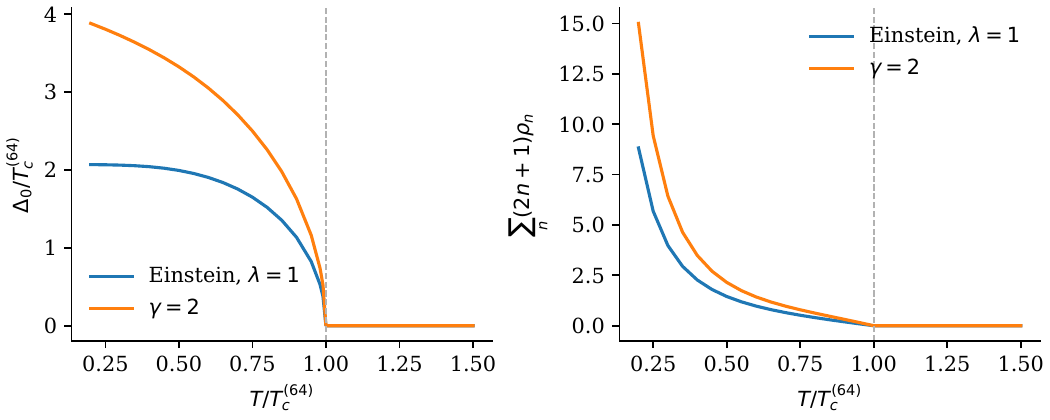}
\caption{Normalized gap $\Delta_0/T_c^{(64)}=\pi(T/T_c^{(64)})\tan\theta^*_0$ at the lowest Matsubara frequency (left) and order parameter $\|\rho^*\|_w=\sum_n(2n+1)\rho^*_n$ (right) of the minimizer, for Einstein phonons with $\lambda=1$ and for the $\gamma$-model with $\gamma=2$, on the truncated chain with $N=64$. The plotted points are discrete illustrative samples; at and above $T_c^{(64)}$ the normal state is shown, whose normalized gap vanishes identically, so the drawn zero is not a limiting value extracted from the superconducting branch. For the plotted truncation, continuous vanishing at $T_c^{(64)}$ follows from the finite-dimensional version of Theorem~\ref{thm:cont} stated in Remark~\ref{rem:finite}; it is not a conclusion drawn from these samples.}\label{fig:gap}
\end{figure}

\begin{figure}[t]
\centering
\includegraphics[width=\textwidth]{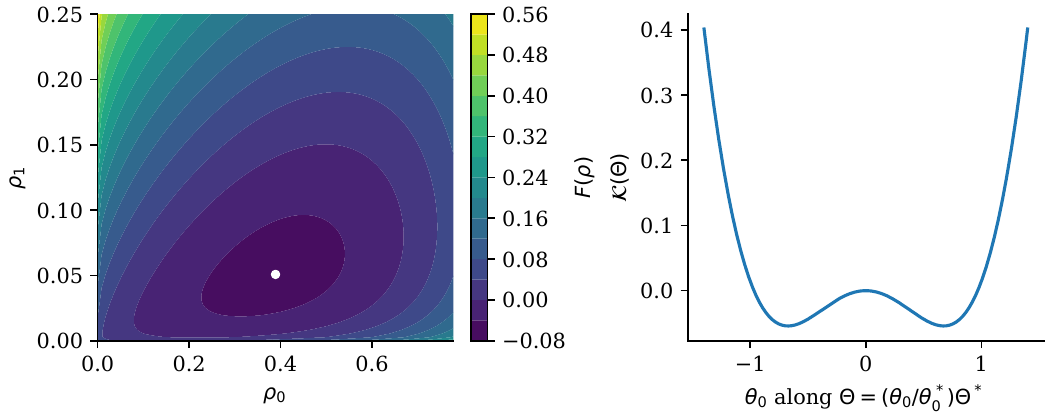}
\caption{Einstein phonons, $\lambda=1$, $T=0.7\,T_c^{(64)}$, $N=64$. Left: $F(\rho)$ as a function of $(\rho_0,\rho_1)$ with the other coordinates fixed at the minimizer (white dot); the level sets are those of a convex function (Proposition~\ref{prop:convex}). Right: $K$ along the ray $t\mapsto t\Theta^*$, parametrized by $\theta_0=t\theta^*_0$: the two minima $\pm\Theta^*$ and the local maximum at the normal state.}\label{fig:landscape}
\end{figure}

Fig.~\ref{fig:gap} shows the normalized gap $\Delta_0/T_c^{(64)}=\omega_0\tan\theta^*_0/T_c^{(64)}$ with $\omega_0=\pi T$, and the order parameter $\|\rho^*\|_w$ across the transition, and Fig.~\ref{fig:landscape} the convexity of $F$ in a two-dimensional slice together with the non-convexity of $K$ along the ray through the minimizer. The script, all data and the figures are in the supplementary material.

\subsection{Use of large language models}\label{sec:ai}

This research used large language models (LLMs) for substantive scientific and editorial work. Under the research brief set by the author, LLM-based agents selected the thermodynamic conjectures, searched the literature, explored proof strategies, drafted the proofs and manuscript, and implemented the numerical study and Lean formalization. OpenAI GPT and DeepSeek models also performed adversarial reviews. Model agreement is not a proof: the arguments are given in the text, and the specified finite-dimensional components are additionally machine-checked in Lean. The author retains responsibility for the final manuscript. No LLM is an author of this paper.

\section{Scope and open problems}\label{sec:scope}

\emph{Coulomb repulsion.} With a finite cutoff, a term $+\mu^*\sum_{0\le n,m<n_c}x_nx_m$ (up to normalization) adds $+\mu^*\sum_{0\le n,m<n_c}\sqrt{\rho_n\rho_m}$ on the first quadrant, where $n_c$ is the cutoff. It replaces $W_{nm}$ by $W_{nm}-2\mu^*$ on that block, so the coefficient signs required by our convexity and folding arguments need not persist. This does not imply failure for every $\mu^*>0$: at fixed interaction and cutoff they persist for sufficiently small $\mu^*$. We prove no general result for Coulomb-repulsive kernels here. Concurrent work states uniqueness at a finite cutoff for sufficiently small $\mu^*>0$~\cite{Siraj2026} and, without spin flips, non-existence of real even-frequency solutions at and above the linear-instability temperature with a symmetric Coulomb term~\cite{GSI2026}.

\emph{The order of the transition.} Theorem~\ref{thm:cont} shows that the transition is continuous: the minimizing density $\rho^*$ is continuous in the weighted norm $\|\cdot\|_w$, hence each fixed-frequency gap $\Delta_n$ is continuous. It establishes no critical exponent and no differentiability of $T\mapsto\rho^*_T$ at $T_c$. Wada concluded, under the assumption of a simple isotope effect, that the Eliashberg transition is of second order~\cite{Wada1964}; our results neither imply nor contradict that conclusion, which rests on additional assumptions. Whether, in the present setting, the mean-field scaling $\|\rho^*_T\|_w\sim(T_c-T)$ holds would require a non-degeneracy condition on the top eigenvector of the stability operator and a local bifurcation analysis; we have not carried this out. Siraj states the lower bound $-C(T_c-T)^2$ on the condensation energy~\cite{Siraj2026}.

\emph{Validity of the theory.} Our theorems concern the Eliashberg functional on its own configuration space. Yuzbashyan and Altshuler showed that in a broader microscopic variational setting, at couplings above a threshold, none of the Eliashberg stationary points is the true global minimum of the free energy~\cite{YuzbashyanAltshuler2022Breakdown}; this is a statement about the validity of the Migdal--Eliashberg approximation, not about the functional studied here.

\emph{Monotonicity of the normalized gap.} The minimizer $\rho^*$ is non-increasing in $n$ in all our numerical examples. Equivalently, $\Delta_n/\omega_n=\sqrt{\rho^*_n/(1-\rho^*_n)}$ is non-increasing, which is the ordering condition used in the companion comparison principle. A proof of this ordering for the nonlinear minimizer would be a first step towards nonlinear comparison results.

\emph{Other solutions.} Theorem~\ref{thm:EL} concerns positive solutions. The sign-changing solutions of the Eliashberg equations~\cite{YuzbashyanAltshuler2022SpinChain,KAY2025II} are critical points of $K$, and by Theorem~\ref{thm:exist} none of them is a global minimizer at any $T>0$. Whether they are saddle points, as found for the $\gamma$-model at $T=0$~\cite{AbanovChubukov2020,ZhangWuAbanovChubukov2022}, or can be local minima, is not addressed here; Siraj states that his results do not classify them either~\cite{Siraj2026}. Complex and non-planar configurations are treated in~\cite{Siraj2026,GSI2026}.

\backmatter

\bmhead{Supplementary information}
The supplementary archive contains the Lean project (source and build instructions) and the script that produces the numerical results and figures of Section~\ref{sec:numerics}, with its output data.

\section*{Statements and Declarations}

\bmhead{Funding}
This work was carried out at, and funded by, SuperMatics Inc. It received no external funding.

\bmhead{Competing interests}
The author is a founder of SuperMatics Inc., which develops computational methods for materials discovery. The author declares no other competing interests.

\bmhead{Ethics approval and consent to participate}
Not applicable.

\bmhead{Consent for publication}
Not applicable.

\bmhead{Data availability}
No experimental data were generated. The supplementary archive contains the program that produces the paper's numerical results, with its output.

\bmhead{Materials availability}
Not applicable.

\bmhead{Code availability}
The Lean formalization and the numerical script are included in the supplementary archive.

\bmhead{Author contribution}
D.R.\ commissioned and directed the research. LLM-based agents selected the target under that brief and generated the initial mathematical arguments, computational implementations and manuscript, as described in Section~\ref{sec:ai}. The author is responsible for the final work.

\bibliography{refs}

\end{document}